\documentclass[5p,times]{elsarticle}
\usepackage{graphicx}
\usepackage{dcolumn}
\usepackage{color}
\usepackage{float}
\usepackage{bm}
\usepackage{epsfig}
\usepackage{graphicx}
\usepackage{psfrag}
\usepackage{amsmath,amssymb}
\usepackage{lineno}
\usepackage{colordvi}
\usepackage{amsfonts}
\usepackage{enumerate}
\usepackage{slashed}
\usepackage{color}
\usepackage{xcolor}
\usepackage{soul}
\usepackage{ulem}
\usepackage{microtype}

\def\gtorder{\mathrel{\raise.3ex\hbox{$>$}\mkern-14mu
 \lower0.6ex\hbox{$\sim$}}}
\def\ltorder{\mathrel{\raise.3ex\hbox{$<$}\mkern-14mu
 \lower0.6ex\hbox{$\sim$}}}
\def\mugegm{\mu_p G_E^p / G_M^p}
\def\gegm{G_E^p / G_M^p}

\def\gep{G_E^p}
\def\gmp{G_M^p}
\def\gen{G_E^n}
\def\gmn{G_M^n}

\newcommand{\order}{ {\cal O} }
\providecommand{\be}{\begin{equation}}
\providecommand{\ee}{\end{equation}}

\allowdisplaybreaks

\begin{document}

\title{Proton and neutron electromagnetic form factors and uncertainties}

\author[anl]{Zhihong Ye}
\ead{yez@anl.gov}
\author[anl]{John Arrington}
\ead{johna@anl.gov}
\author[pitp,fermi,uk]{Richard J. Hill}
\ead{rjh@fnal.gov}
\author[iit,cor,kor]{Gabriel Lee}
\ead{leeg@physics.technion.ac.il}

\address[anl]{Physics Division, Argonne National Laboratory, Argonne, Illinois 60439}
\address[pitp]{Perimeter Institute for Theoretical Physics, Waterloo, ON, N2L 2Y5 Canada}
\address[fermi]{Fermilab, Batavia, IL 60510, USA}
\address[uk]{Department of Physics and Astronomy, University of Kentucky, Lexington, KY 40506, USA}
\address[iit]{Physics Department, Technion---Israel Institute of Technology, Haifa 32000, Israel}
\address[cor]{Department of Physics, LEPP, Cornell University, Ithaca, NY 14853, USA}
\address[kor]{Department of Physics, Korea University, Seoul 136-713, Republic of Korea}

\date{July 27, 2017}

\begin{abstract}
We determine the nucleon electromagnetic form factors and their uncertainties
from world electron scattering data. The analysis incorporates 
two-photon exchange corrections, constraints on the low-$Q^2$ and high-$Q^2$ behavior,
and additional uncertainties to
account for tensions between different data sets and uncertainties in radiative corrections.
\end{abstract}

\begin{keyword}
Elastic Scattering;
Form Factors \\
{\it Preprint}: FERMILAB-PUB-17-281-T
\PACS 25.30.Bf, 13.40.Gp, 14.20.Dh
\end{keyword}

\maketitle


\section{Introduction}

The proton and neutron electromagnetic form factors are precisely
defined quantities encoding the charge and magnetization distributions
within the nucleon. Since the 1950s, these form factors have been
extensively measured using electron scattering.  A new generation of
experiments, frequently utilizing polarization degrees of freedom,
have provided a dramatic increase in our understanding of the form
factors in the last 20 years~\cite{arrington07a, perdrisat07,
  arrington11a, punjabi15}. With the extended $Q^2$ range and improved
precision, these measurements also demonstrated the importance of
two-photon exchange (TPE) effects~\cite{carlson07, arrington11c,
  bernauer11, arrington11b}.

Besides the direct determination of nucleon structure, these form
factors are key inputs to other studies and searches in particle, nuclear, and atomic physics. 
For example, precise
knowledge of neutrino-nucleus interaction cross sections is required
in order to access fundamental neutrino properties at long-baseline
oscillation experiments~\cite{Mosel:2016cwa,Katori:2016yel,Alvarez-Ruso:2017oui};
the electroweak vector form factors of the nucleons
are an important input to these cross sections, and
are determined by an isospin rotation of
the electromagnetic form factors.
Measurements of nuclear structure using the $A(e,e'p)$ reaction
require reliable knowledge of the elastic electron-proton ($ep$)
scattering cross section, as do Coulomb Sum Rule~\cite{meziani84,morgenstern01} 
studies using inclusive quasielastic scattering and
exclusive high-$Q^2$ proton knockout studies of Color
Transparency~\cite{oneill95, abbott98, garrow02, dutta03}.
Other applications include
the determination of fundamental constants from (muonic) atom
spectroscopy~\cite{Mohr:2015ccw}, searches for new particles in
photon-initiated high-energy collider
processes~\cite{Manohar:2016nzj}, and constraints on QCD chiral
structure and new forces in parity-violating electron-proton
scattering~\cite{armstrong05, beise05, Androic:2009aa, armstrong12}. 
The impact of TPE on some of these observables is discussed in Refs.
~\cite{arrington04a, arrington07b, carlson07, arrington11b}.

Recent high-precision form factor measurements, coupled with our new
understanding of the importance of TPE contributions and the need for
reliable uncertainty estimates on a range of important derived
observables, call for an updated global analysis of the nucleon
form factors. Several commonly-used parameterizations have one or
more limitations. The Bosted~\cite{bosted94} parameterization was
generated before the polarization data were available and does not
include any correction for TPE, although this fit and the
TPE-uncorrected results from Refs.~\cite{arrington07c, arrington07b} are still useful
parameterizations of $ep$ cross sections, with the TPE contribution
absorbed into effective proton form factors. The fits by
Brash~\cite{brash02}, Kelly~\cite{kelly04}, Graczyk~\cite{graczyk10} and  
Sufian~\cite{sufian17} include a mix of cross-section and
polarization data, but without the TPE corrections necessary to yield
consistent results. Fits by Alberico~\cite{alberico09} and
Qattan~\cite{qattan12, qattan15} include phenomenological TPE corrections extracted
from the difference between Rosenbluth and polarization measurements, but these
extractions require assumptions about $\varepsilon$ and $Q^2$ dependence, and
the data do not provide significant constraints on the corrections at low 
$Q^2$. Finally, several works~\cite{brash02, arrington04b, 
arrington07c, bernauer14} only provide fits to proton data while
others~\cite{lomon01, lomon02, budd03, budd04, arrington04b,
  bradford06, arrington07c, bodek08, qattan11, qattan12} do not
provide uncertainties. References~\cite{arrington07b} and~\cite{venkat11}
provide relatively complete analyses, but the former focused on the
low-$Q^2$ region (below 1~GeV$^2$) and the latter evaluates, but does
not provide, a parameterization of the uncertainties. Many of these form factor
parameterizations are sufficient for specific purposes or in limited
kinematic regimes, but the experimental progress and improved
understanding of TPE call for a more complete analysis.

The goal of this work is to provide a parameterization of proton and
neutron electromagnetic form factors and uncertainties using the complete world data set
for electron scattering,
and applying our best knowledge of the TPE corrections. Additional
systematic errors are included to account for estimated uncertainties in
TPE and tensions between data sets. We aim to provide a reliable
parameterization covering both low-$Q^2$ and high-$Q^2$ regions, with
sufficiently conservative errors such that it is safe to use
these form factors as input to calculations or analyses that need to
represent the present state of uncertainties.
Where significant ambiguities exist, e.g., in the
choice of external constraints on the proton charge radius, 
separate fits can be used to estimate the sensitivity of derived
observables to data selections. 
In forthcoming work we will examine illustrative applications and 
a range of fits making specific assumptions
about the proton radius and the choice of data sets~\cite{nextpaper17}.


\section{Definitions and notation}

The cross section for electron-nucleon scattering in the single-photon exchange approximation
can be expressed in terms of the Sachs form factors $G_E^N$ and $G_M^N$ as
\begin{align}
\left( \frac{d\sigma}{d \Omega} \right)_{0} = \left(
\frac{d\sigma}{d\Omega} \right)_{\rm Mott} \frac{\epsilon (G_E^N)^2 + \tau
  (G_M^N)^2}{\epsilon (1+\tau)} \,,
\label{eqn:xs1photon}
\end{align}
where $N=p$ for a proton and $N=n$ for a neutron, 
$(d\sigma/d\Omega)_{\rm Mott}$ is the recoil-corrected relativistic point-particle (Mott) cross section,
and $\tau$, $\epsilon$ are dimensionless kinematic variables:
\begin{align}
\tau = \frac{Q^2}{4m_N^2} \,, \quad
\epsilon = \left[ 1 + 2 (1 + \tau) \tan^2 \frac{\theta}{2} \right]^{-1} \,,
\end{align}
with $\theta$ the angle of the final state electron with respect to the incident beam direction and 
$Q^2 = -q^2$ the negative of the square of the four-momentum transfer $q$ to the nucleon.

Radiative corrections modify the cross section: 
\begin{align}\label{eq:radcor}
  d\sigma = d\sigma_0 (1+\delta) \,,
\end{align}
where $d\sigma_0$ is the Born cross section in Eq.~(\ref{eqn:xs1photon}).%
\footnote{
  The form factors are interpreted in the renormalization scheme
  defined in Ref.~\cite{lee15}, which is a simplification of Ref.~\cite{maximon00}.
  The $ep$ cross sections presented in Sec.~\ref{sec:elastic} are interpreted
  using the Maximon-Tjon convention~\cite{maximon00} for soft photon subtraction. 
  The relation of these conventions to a standard minimal subtraction
  (${\rm MS}$) factorization scheme is given in Ref.~\cite{Hill:2016gdf}.
  }
Radiative corrections were already applied to the published
cross sections we include in this fit, but we apply additional TPE
corrections and modify the corrections applied for some experiments, as
described in the following section.


\section{Data sets and corrections}

This section provides an overview of our data selections and applied corrections.  
We discuss separately the proton and neutron data sets. 


\subsection{Proton data}

For the proton, we fit directly to unpolarized cross section data~\cite{dudelzakphd, janssens66,
bartel66, albrecht66, frerejacque66, albrecht67, goitein67, litt70, goitein70, berger71, price71, ganichot72,
bartel73, kirk73, borkowski74, murphy74, borkowski75, stein75, simon80, simon81, bosted90, rock92, sill93,
walker94, andivahis94, dutta03, christy04, qattan05, bernauer14} and to $\gegm$ ratios extracted from
polarization data~\cite{milbrath99, pospischil01, gayou01, strauch02, punjabi05, maclachlan06,
jones06, crawford07, ron11, zhan11, paolone10, puckett12, puckett17}. Note that the data taken from
Refs.~\cite{punjabi05, puckett12, puckett17} include updated extractions of $\gegm$ from 
Refs.~\cite{jones00,gayou02,puckett10,meziane11}, and we use these updated extractions in our analysis.
Following the procedures described in Refs.~\cite{arrington03a, arrington04a}, we apply updated
radiative corrections to several of the older measurements, exclude the small-angle data
from Ref.~\cite{walker94}, and split up data sets~\cite{goitein70, bartel73, andivahis94}
taken under different conditions into two or more
subsets with separate normalization factors.

After examining the systematic uncertainties in each of these experiments, we implement some adjustments to
make the assumptions more consistent (e.g., uncertainties associated with TPE) or to ensure that
the uncertainties were separated into uncorrelated and normalization factors in a consistent 
fashion. In Refs.~\cite{albrecht67, price71, bartel73} and~\cite{goitein70} (back-angle data), the
common systematic uncertainties were included in the point-to-point systematics. We remove
these common systematics from the point-to-point contributions and apply them instead as
additional contributions to the normalization uncertainty. To make the uncertainties applied for
radiative corrections more consistent across experiments,
we increase the normalization uncertainty in Refs.~\cite{simon80, simon81} from
$\sim$0.5\% to 1.5\% and add 0.5\% in quadrature to the point-to-point uncertainty to account
for the use of older radiative correction procedures and the neglect of uncertainty associated with
TPE corrections. We add a 1\% point-to-point uncertainty to the data from Ref.~\cite{murphy74} to
be more consistent in estimating the uncertainties from radiative corrections.
In Ref.~\cite{qattan05}, uncertainties were separated into normalization,
point-to-point, and ``slope'' uncertainties, i.e., correlated systematics that varied linearly with
$\varepsilon$, to maximize sensitivity to deviations from a linear $\varepsilon$ dependence. 
To make this data set consistent with other world data, we replace the slope uncertainty with an
additional point-to-point systematic (0.32\%, 0.28\%, and 0.22\% for $Q^2$ = 2.64, 3.2, and 4.1~GeV$^2$,
respectively), such that the total uncertainty on $\mugegm$ matches the original extraction including
both point-to-point and slope uncertainties.

For the new data from the A1 collaboration~\cite{bernauer14}, we use the rebinned data with additional systematic
uncertainties as provided in the Supplemental Material of Ref.~\cite{lee15}. In addition,
because Ref.~\cite{bernauer14} also quotes correlated systematic uncertainties modeled as
cross-section corrections that vary linearly with the scattering angle, we use the procedure
described in Ref.~\cite{lee15} and take the coefficients of the $\theta$-dependent corrections as
additional fit parameters 
(similar to the normalization uncertainties applied to the different data subsets), 
so that the full uncertainties from all data sets are included in the fit.%
\footnote{The procedure is described in Section VI.C.3 of Ref.~\cite{lee15} and
  is represented by the line ``Alternate approach'' in Table XIV.}

For all cross-section measurements, TPE corrections are applied as described in Ref.~\cite{lee15} using
the ``SIFF Blunden'' calculation following the prescription of Ref.~\cite{blunden05a}.%
\footnote{
  As discussed in Refs.~\cite{Hill:2016gdf, lee15}, the hard TPE corrections depend
  on the scheme used to apply radiative corrections to the data, typically based
  on either Refs.~\cite{mo69} or \cite{maximon00}. 
  These small differences, as well as differences in hadronic vacuum polarization corrections
  and in higher-order radiative corrections, 
  are absorbed into the radiative correction uncertainty budget.
}
The uncertainties included for radiative corrections in the cross-section data are assumed to be
sufficient to cover TPE uncertainties at low $Q^2$ after application of the SIFF correction described
above. At larger $Q^2$, the hadronic calculations are not expected to be as reliable and we include
an additional uncertainty based on the analysis of Ref.~\cite{arrington07c}.
In this work, the following additional correction is applied at high $Q^2$ to resolve the small remaining 
difference between polarization data and TPE-corrected Rosenbluth extractions:
\begin{equation}
\delta_{2\gamma} \to \delta_{2\gamma} + 
  0.01 ~ [\varepsilon-1] ~ \frac{\ln{Q^2}} {\ln{2.2}}~~~~~(Q^2>1~{\rm GeV}^2) \,,
\label{eq:extratpe}
\end{equation}
where $\delta_{2\gamma}$ is the contribution of TPE to the radiative correction in Eq.~(\ref{eq:radcor}).
This has minimal impact on the final extraction of $\mugegm$, but is important in
the extraction of $\gmp$ at high $Q^2$. We use this purely phenomenological additional correction
to estimate potential systematic uncertainties to the high-$Q^2$ extractions. We perform the global fit
with and without this extra correction and take the difference as a systematic uncertainty on the final form
factors. Note that the additional correction is
always negative, and increases the Born cross section inferred from data according to Eq.~(\ref{eq:radcor}). 

While recent comparisons of positron and electron scattering~\cite{adikaram15, rachek15, rimal17,
henderson17}  support the idea that TPE yields an angle-dependent correction to the cross sections
that may explain the discrepancy between cross-section and polarization data, we do not yet have
precise measurements of the correction. For this analysis, we assume that after applying the TPE
contributions based on Ref.~\cite{blunden05a}, the remaining uncertainty is accounted for in the
radiative correction uncertainties applied to the individual data sets
(typically a combination of uncorrelated and normalization factors). 
As in previous analyses~\cite{arrington07b, arrington07c}, 
we do not apply TPE corrections to the polarization data.
As discussed in these works, the estimated corrections are small compared to the experimental
uncertainties, even accounting for significant uncertainty in the calculations~\cite{meziane11,
borisyuk11, guttmann11} and the fact that this is a correlated correction across all polarization
measurements.

The updated proton data set used in our fit is included in the Supplemental
Material~\cite{supplemental}.


\subsection{Neutron data} \label{sec:ndata}

For the neutron, we perform separate fits to the charge and magnetic form factor data.
Many early attempts to extract neutron form factors involved cross-section
measurements on the deuteron ($d$), where isolating the
neutron contribution involved subtracting the dominant proton contribution after
accounting for nuclear effects in the deuteron.
Such extractions involve large corrections for final-state interactions and other effects. 
Later measurements, using polarization degrees of freedom or ratios of proton knockout to neutron
knockout cross sections, 
typically have much smaller corrections and are thus more reliable. For both $\gen$ and $\gmn$, 
we select experiments that had minimal corrections and model-dependent uncertainties in their range of $Q^2$.
In some cases, we make adjustments such that the quoted errors are more complete
and consistent between different data sets, as we now describe.

The updated $\gen$ and $\gmn$ data sets used in our fit is included in the Supplemental Material~\cite{supplemental}.

\subsubsection{\texorpdfstring{$G_M^n$}{Lg} data}

For $\gmn$, we take data from Refs.~\cite{rock82, lung93, gao94, anklin98, kubon02,
anderson07, lachniet09}. Even with this limited data set of more reliable extractions, there is
tension between the data as published. After examining the experiments more carefully, we make
some modifications for corrections or uncertainties that were not fully
accounted for in the original works. These modifications are as follows. 

For Ref.~\cite{rock82}, a later analysis~\cite{rock92} provided updated values of the ratio
$\sigma_n/\sigma_p$, but not updated $\gmn$ values. We correct the quoted $\gmn$ values from the
original publications to account for the updated $\sigma_n/\sigma_p$ analysis, and apply a
correction (from 0.6--1.4\% on $\gmn$) to account for the fact that the original analysis assumed
$\gen=0$. We also apply an additional 0.5\% to the $\gmn$ uncertainties to better account for the
uncertainty in the $ep$ cross section used in the original result, and a 1\%
normalization uncertainty for this data set (as well as for Ref.~\cite{lung93}) to account for
correlated uncertainties associated with the use of older estimates for radiative corrections
and model dependence. Other experiments are assumed to have a 0.5\% normalization uncertainty.

For Refs.~\cite{anklin98, kubon02}, older parameterizations were used in determining the $ep$ 
cross section and the $\gen$ contribution to the $en$ cross section. We make updated estimates
of the uncertainties based on the difference in the corrections and uncertainties 
applied in the original work and in more recent form factor evaluations. 

The results of Ref.~\cite{lachniet09} were generally dominated by systematic uncertainties 
that are likely to have significant correlation between points close together in $Q^2$. 
To better reflect this, we rebin the $\gmn$ points, combining three points for each new $Q^2$ value 
(two points in the highest-$Q^2$ bin); statistical uncertainties are combined in quadrature, but the 
systematic uncertainties are taken as the average of the (nearly identical) systematic uncertainties of the three individual points.

\subsubsection{\texorpdfstring{$G_E^n$}{Lg} data}

The analysis of $\gen$ is based on data from Refs.~\cite{meyerhoff94, eden94, passchier99, herberg99, rohe99,
golak01, schiavilla01, zhu01, bermuth03, madey03, warren04, glazier05, geis08, riordan10, schlimme13}. In most
cases, these measurements used polarization observables that are sensitive only to the ratio
$\gen/\gmn$. Different values and uncertainties for $\gmn$ were used to convert these ratio 
measurements into values for $\gen$, potentially underestimating the uncertainties of the $\gen$
extractions. However, the final $\gen$ uncertainties are large, typically 15\% or more.
Updating all of these extractions to use the same parameterization of $\gmn$ and its uncertainties
would have minimal impact: $\gmn$ is within 5\% of the dipole form for the full $Q^2$ range of $\gen$
measurements, and the differences between different $\gmn$ values used is even smaller. Thus, no
additional uncertainty or correction is applied.

Elastic $e d$ scattering can also be used to extract $\gen$, but there is significant model dependence 
in the result which tends to be nearly identical for different data sets. 
Therefore, we include only one extraction of $\gen$ from $e d$ elastic scattering: 
the analysis of Ref.~\cite{schiavilla01}, which included a detailed estimate of the model dependence.


\section{Global fit procedure}

The fitting procedure follows the general approach of Ref.~\cite{lee15}. For the proton form
factors, we perform a simultaneous fit of $\gep$ and $\gmp$ to the cross-section and polarization
data. For the neutron, we perform separate fits of $\gen$ and $\gmn$ to the extractions of the
individual form factors. In all cases, the fit is a bounded polynomial
$z$-expansion~\cite{hill10}, 
\begin{equation}\label{eq:zexp}
  G(Q^2) = \sum_{k=0}^{k_{\rm max}} a_k z^k \,, \quad
  z = \frac{\sqrt{t_{\rm cut} + Q^2} - \sqrt{t_{\rm cut} - t_0}}{\sqrt{t_{\rm cut} + Q^2} + \sqrt{t_{\rm
cut} - t_0} } \,,
\end{equation}
where $G$ stands for $G_E^p$, $G_E^n$, $G_M^p/\mu_p$ or $G_M^n/\mu_n$, 
and $t_{\rm cut}=4 m_\pi^2$.%
\footnote{
For $\gmn$, the normalization at $Q^2=0$ is the numerical value of the neutron magnetic moment when expressed in units of $e/2m_n$. If one uses the value of $\mu_n$ in nuclear magnetons ($e/2m_p$) or if the form factors are defined in a different convention, e.g., using an average nucleon mass for both the proton and neutron, there will be differences between the data sets at the level of the proton-neutron mass difference. These differences are negligible compared to other sources of normalization uncertainty, and so will be corrected for when fitting the normalization factor for each experiment. The same is true for other similar approximations, e.g., the use of the proton mass or an average nucleon mass in defining $\tau$ for the neutron.
}
We choose a fixed value of $t_0=-0.7$~GeV$^2$ for all four
form factors so that there is a single definition of $z$ in all cases.  The value
$t_0 = -0.7$~GeV$^2$ is a compromise between the broad $Q^2$ range for
proton cross-section data ($Q^2 \sim 0$--$30$~GeV$^2$) and the limited $Q^2$ range for $\gen$ form factor data ($Q^2 \sim 0$--$3.5$~GeV$^2$).

Sum-rule constraints are applied on each form factor to ensure appropriate
behavior in the limits of small and large $Q^2$.
One sum rule is applied to enforce the correct normalization at $Q^2=0$.
Four additional sum rules ensure the asymptotic scaling 
$G \sim Q^{-4}$ at large $Q^2$; i.e., $Q^i G(Q^2) \to 0$ as $Q^2 \to \infty$ ($z \to 1$)
for $i=0\dots 3$.
With these five sum rules in place, the number of free parameters is $k_{\rm max} -4$. 
Following Ref.~\cite{lee15}, bounds are
applied to the coefficients $a_k$ using a normalized Gaussian
prior $\left| {a_k}\right| < 5$.

With the bounds on the coefficients in place, 
we can add an arbitrary number of fit parameters, i.e., varying $k_{\rm max}$ in Eq.~(\ref{eq:zexp}), 
without the fit uncertainties growing out of control.
Thus, while good fits are obtained with $k_{\rm max}=10$ for the proton and $k_{\rm max}=7$
(10) for $\gen$ ($\gmn$), we
perform the proton fits with $k_{\rm max}=12$ and neutron fits with $k_{\rm max}=10$. 
This ensures that the fit is not strongly influenced by the $k_{\rm max}$ truncation,
while retaining a manageable number of independent fit parameters.  

When extrapolating to larger $Q^2$, the form factors are influenced by
higher-order parameters that are not directly constrained by data.
We include high-$Q^2$ ``constraint'' points as theoretical priors to avoid a sudden and dramatic 
increase or decrease of the form factors when going beyond the range of the data.
These are listed in the Supplemental Material~\cite{supplemental}.

Tensions between different electron-nucleon scattering data sets and between low-$Q^2$ and high-$Q^2$
data~\cite{lee15} suggest that a global fit to all data, up to $Q^2 \approx 30$~GeV$^2$,
may not yield the most reliable result for the charge and magnetic radii.
Rather than allowing the radii to float in the fit, we constrain them from external measurements,
or fix them to ``consensus'' values obtained from dedicated analyses specifically aimed at isolating
the radii.

For the neutron electric radius, we include the precise value from neutron-electron scattering length
measurements, $(r_E^n)^2 = -0.1161(22)$~fm$^2$~\cite{pdg16}, as a data point in the fit.
A precise value of the proton electric radius, $r_E^p$, has been extracted from muonic hydrogen Lamb shift
spectroscopy~\cite{antognini13}.  However, given the unresolved
status of the proton radius puzzle~\cite{pohl13,carlson15,Hill:2017wzi}, we
do not include this point in our fit.  We take instead the CODATA consensus central value
$r_E^p=0.879$~fm~\cite{Mohr:2015ccw} based only on $ep$ scattering results~\cite{arrington15a}. 
For the magnetic radii we take PDG consensus central values~\cite{pdg16}, $r_M^n = 0.864$~fm
and $r_M^p = 0.851$~fm.%
\footnote{
  For $r_M^p$, we use the average of the Mainz and world values presented in Ref.~\cite{lee15},
  whereas Ref.~\cite{pdg16} adopts the Mainz value.  
}
For $r_E^p$, $r_M^p$ and $r_M^n$, we force the fit to reproduce the consensus central
value, but release the radius constraints when evaluating the fit uncertainty.
These fits should not be interpreted as providing new information on the nucleon electromagnetic radii,
but are designed to summarize the implications of world scattering data for form factors and
uncertainties throughout the entire $Q^2$ range. 


\begin{figure*}[ht!]
\begin{center}
\includegraphics[angle=0,width=0.48\textwidth,height=8.0cm]{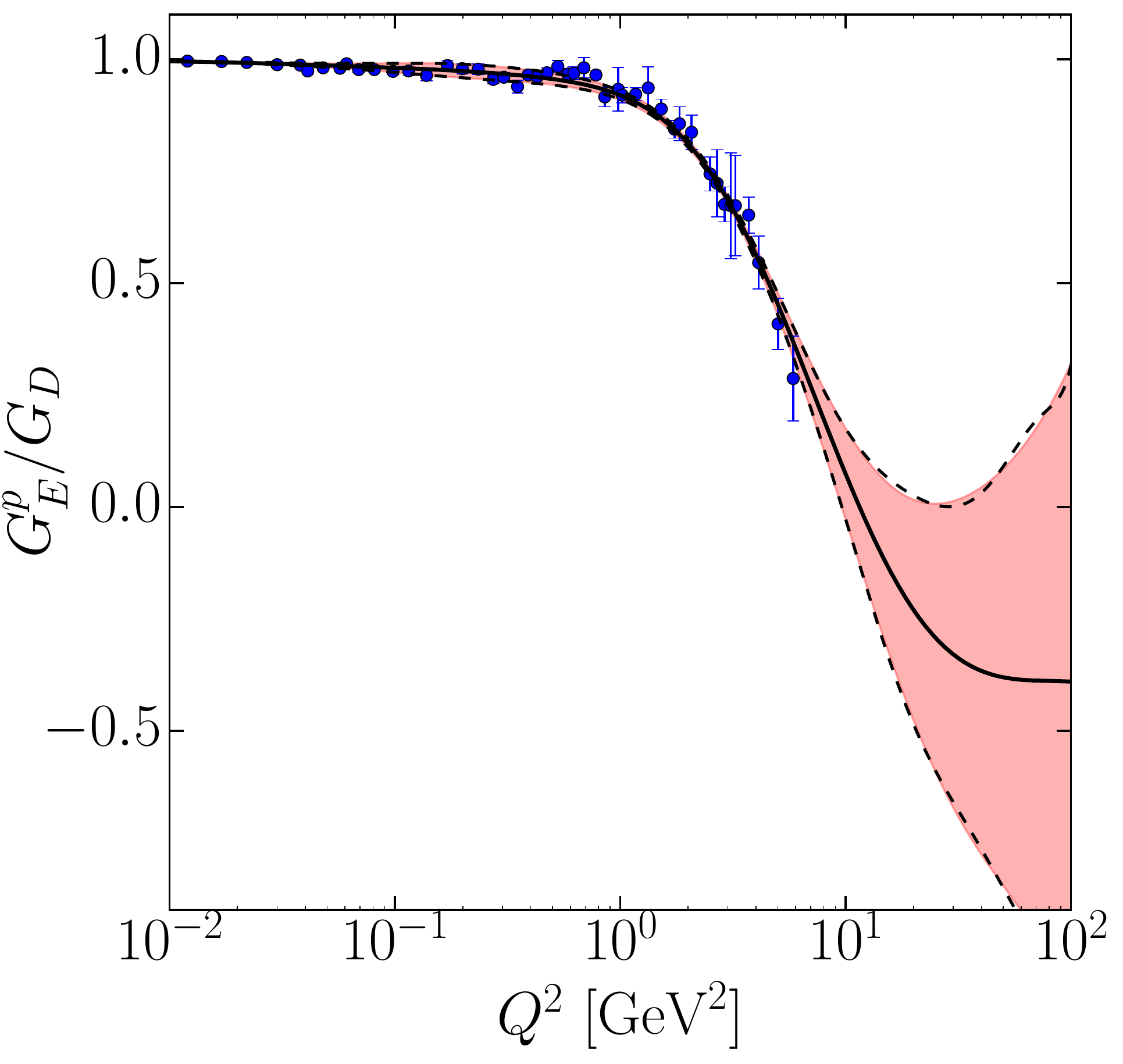}
\includegraphics[angle=0,width=0.48\textwidth,height=8.0cm]{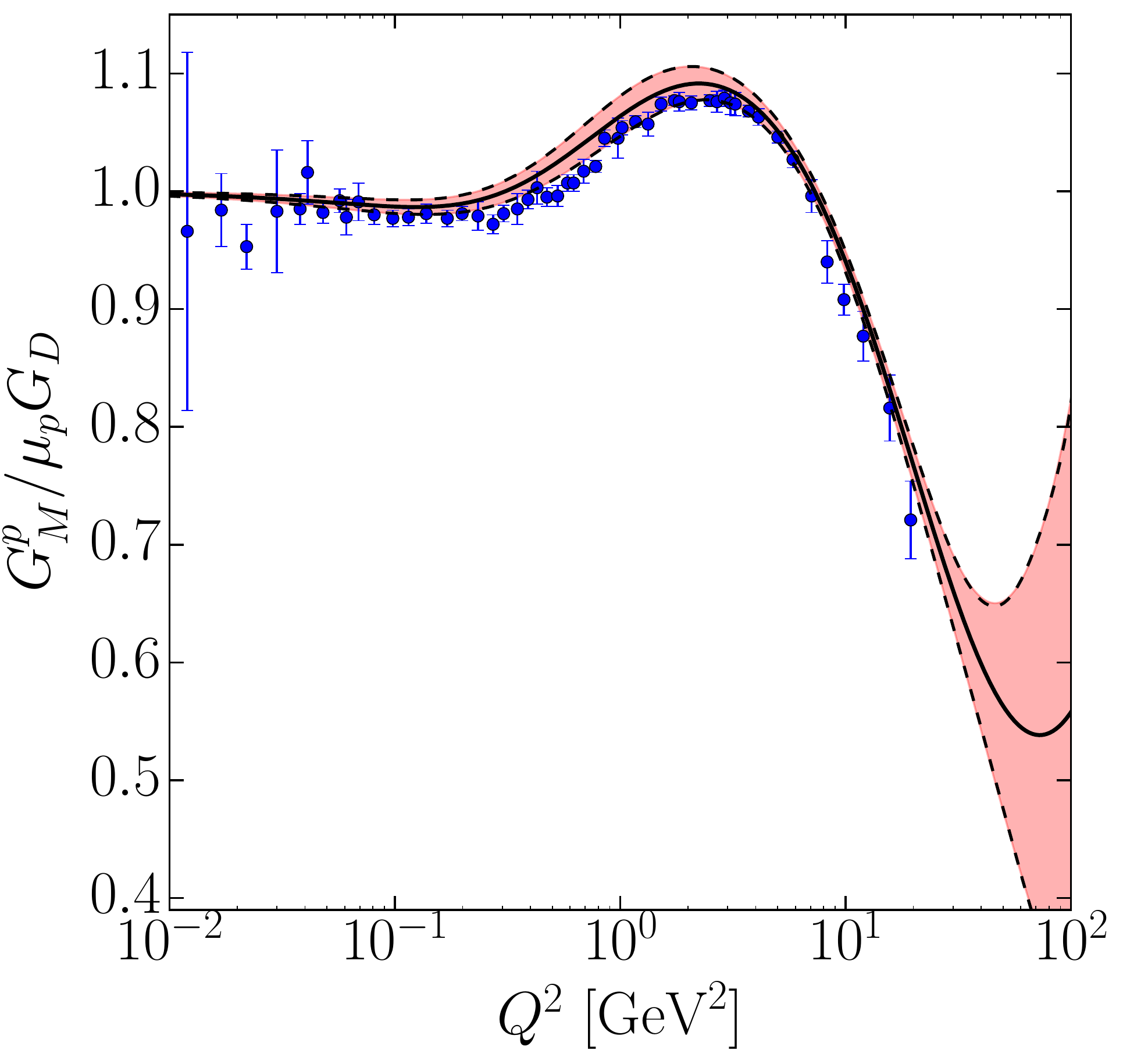}
\caption{(Color online) Parameterization of $\gep/G_D$ (left) and $\gmp/\mu_pG_D$ (right) from the global fit of
proton cross-section and polarization data (solid curves). The red shaded band indicates the total uncertainty,
including the fit uncertainty from the error matrix and additional systematic uncertainties described in the text
and shown in Fig.~\ref{fig:protonuncertainties}.
The dashed curves are the parameterizations of the total uncertainty bands (provided in the Supplemental Material).
The blue circles are taken from the 2007 global analysis of Ref.~\cite{arrington07c} to provide a comparison to
direct LT separations  from a previous global analysis and to indicate the kinematic coverage of the world data. 
The new fit yields systematically larger values for $\gmp$ up to $Q^2 \approx 1$~GeV$^2$ because the Mainz
data~\cite{bernauer14}, not included in the fit of~\cite{arrington07c}, yields larger values of $\gmp$ below
1 GeV$^2$, and so increases the normalization of the world data relative to the fit of~\cite{arrington07c}.}
\label{fig:protonfits}
\end{center}
\end{figure*}


For the proton fit, the $\chi^2$ that is minimized is:
\begin{equation}
\chi^2_p = \chi^2_{\sigma} + \chi^2_{\rm ratio} + \chi^2_{\rm norm} + \chi^2_{\rm slope} + \chi^2_{\rm bound} + \chi^2_{\rm radius},
\end{equation}
with contributions from the cross-section and polarization $\gep/\gmp$ ratio data,
normalization parameters for all data sets, slope parameters for Ref.~\cite{bernauer14}
(as detailed in Ref.~\cite{lee15}),
coefficient bounds, and external radius constraints.
For the neutron case, we fit directly to the extracted form factors and the
$\chi^2$ contributions are:
\begin{equation}
\chi^2_n = \chi^2_{\rm ff} + \chi^2_{\rm norm} + \chi^2_{\rm bound} + \chi^2_{\rm radius}.
\end{equation}

Uncertainties are evaluated from the covariance matrix of the fit supplemented by additional
systematic uncertainties. As noted in Ref.~\cite{lee15}, there is a 
tension between the Mainz data~\cite{bernauer14} and other world data, and we include an additional
systematic to account for this. At low $Q^2$, we can directly compare the fits to Mainz and world data to 
estimate this systematic uncertainty, but because the Mainz data are limited to $Q^2 < 1$~GeV$^2$,
the fits diverge rapidly at higher $Q^2$ values. Thus, we take the difference between the fits to the
world (excluding Mainz) and world$+$Mainz data, which becomes small at large $Q^2$ values where
the Mainz data does not contribute.

As noted above, we use the additional TPE contribution at large $Q^2$ values from Ref.~\cite{arrington07c},
Eq.~(\ref{eq:extratpe}), to estimate the high-$Q^2$ TPE uncertainty. Rather
than applying this as an independent uncertainty on each cross-section point, we
estimate the uncertainty by performing the final fit with and without this additional TPE
correction and take the difference in the fits as the systematic uncertainty.

To test for any systematic bias from theoretical priors, we compare the default fit to fits with
different $t_0$ values, with different $k_{\rm max}$, and without the radius or high-$Q^2$
constraints.  The choice
$t_0=t_0^{\rm opt}=t_{\rm cut}\left(1-\sqrt{1+Q^2_{\rm max}/t_{\rm cut}}\right)$,
instead of the default $t_0=-0.7$~GeV$^2$, yielded negligible differences throughout the $Q^2$ range of the data.%
\footnote{
  The ``optimal'' choice of $t_0$ minimizes the maximum size of $|z|$
  in the range $0<Q^2<Q^2_{\rm max}$, with $Q^2_{\rm max}$ equal to the maximum $Q^2$ in a given data set.  
}
Fits with $k_{\rm max}=20$, instead of the default $k_{\rm max}=12$ $(10)$ for the proton (neutron) data,
also show very good agreement with the default fit: the only significant
differences occur at $Q^2$ values above the range of data, where the $k_{\rm max}=20$ fits show somewhat
different behavior and larger uncertainties.
Finally, fits excluding the radius and/or high-$Q^2$ constraints
differ negligibly from the default fit in regions where sufficient data
exist to directly constrain the form factors.


\section{Global fit results}


\begin{figure}[hb!]
\begin{center}
\includegraphics[angle=0,width=0.48\textwidth]{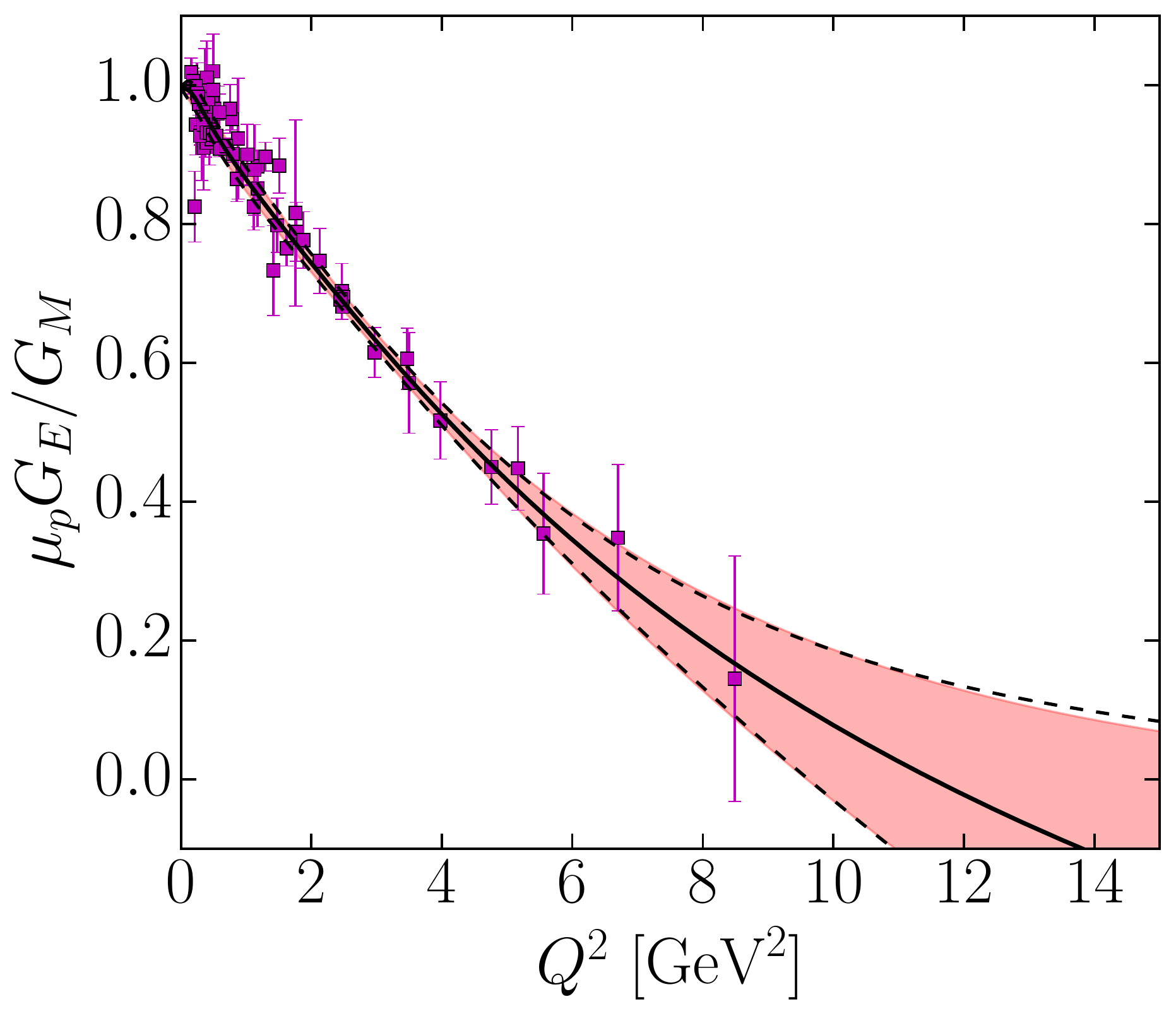}
\caption{(Color online) Parameterization of $\mugegm$ from the global fit of proton data. The error bands are
the same as in Fig.~\ref{fig:protonfits} and the magenta squares are the direct extractions from polarization
measurements.}
\label{fig:protongegm}
\end{center}
\end{figure}



\begin{figure*}[ht!]
\begin{center}
\includegraphics[angle=0,width=0.48\textwidth]{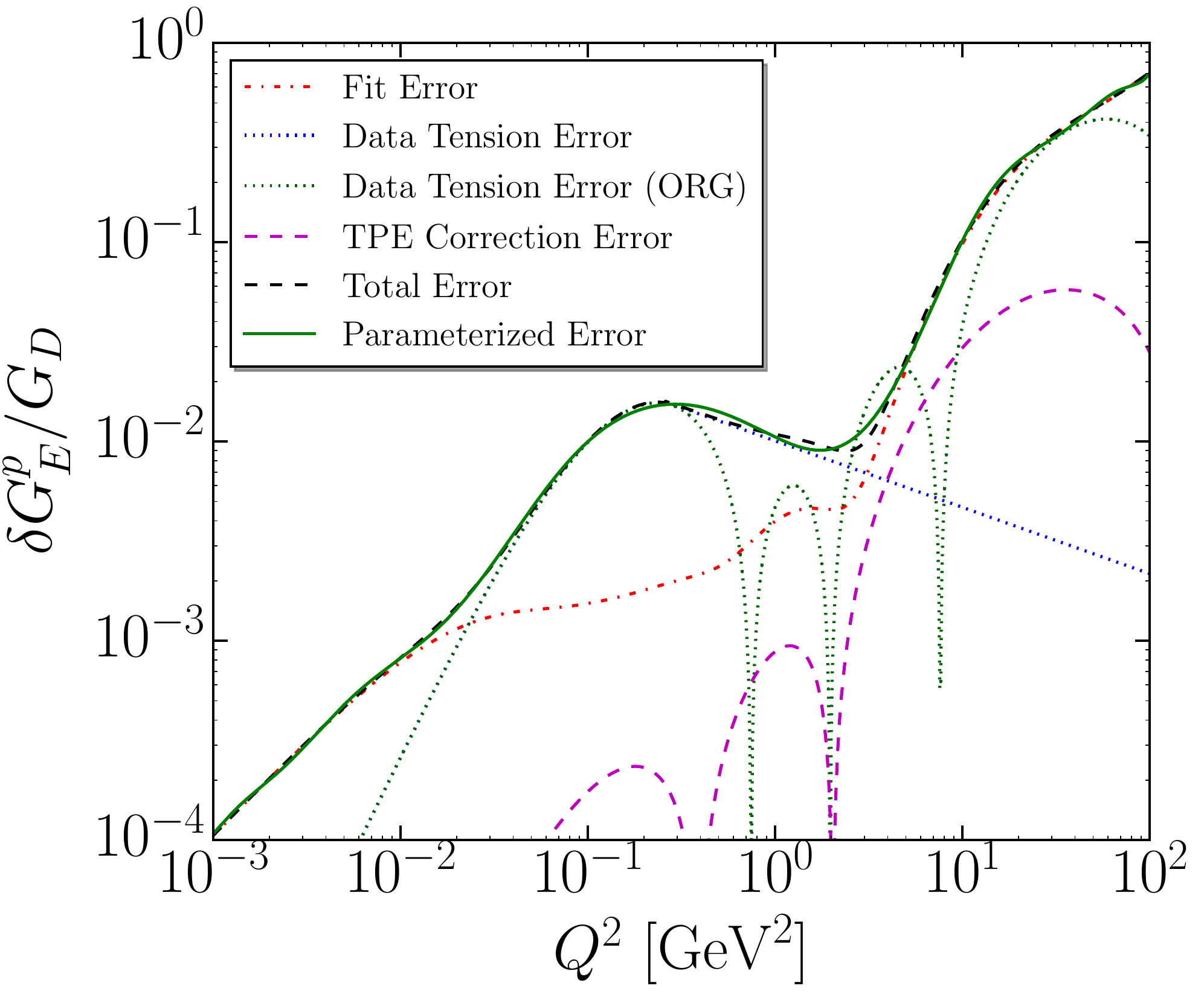}
\includegraphics[angle=0,width=0.48\textwidth]{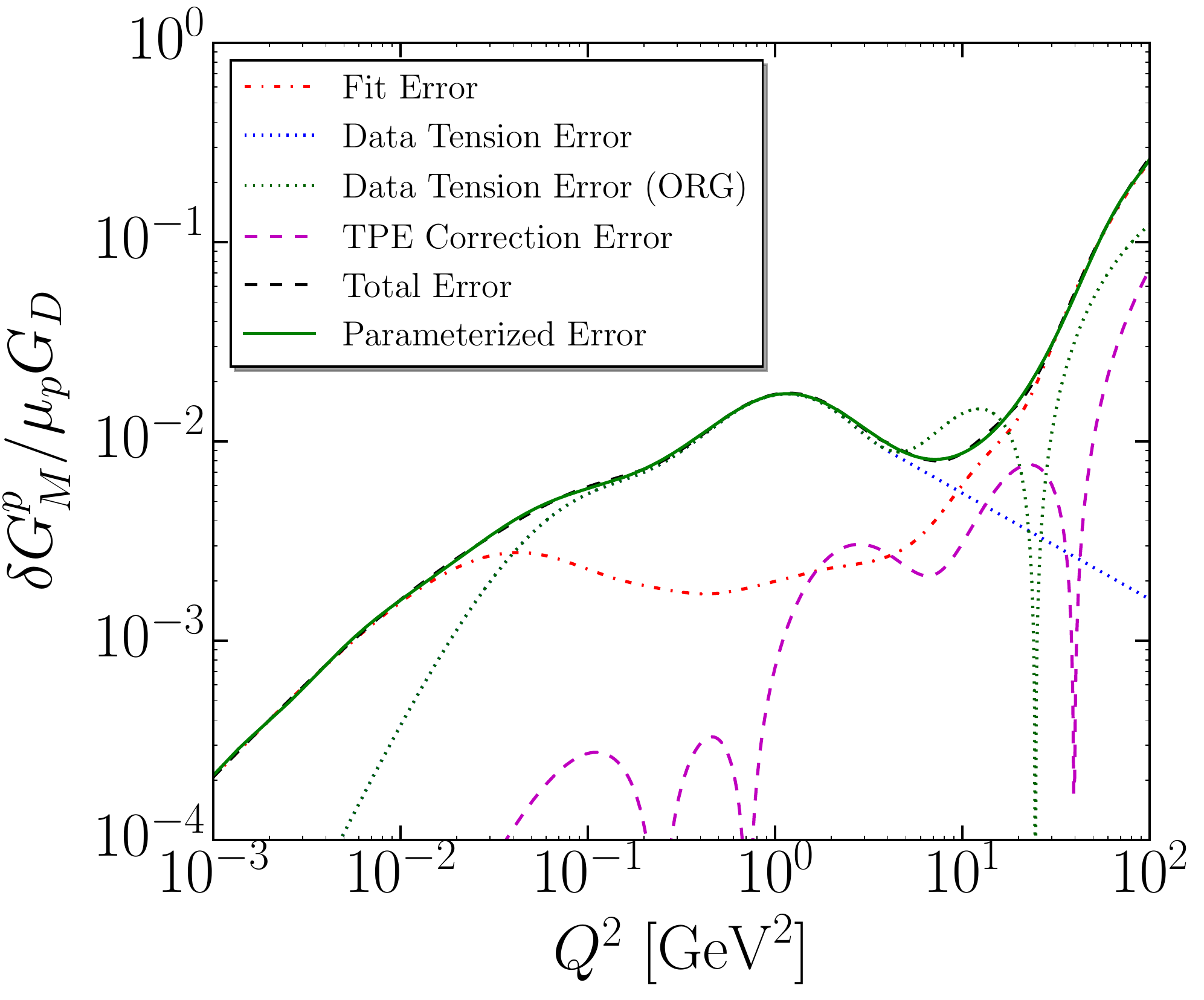}
\caption{(Color online) Contributions to the proton fit uncertainties. 
The red dot-dashed curves are the uncertainties from the fit based on the statistical and systematic uncertainties
of the data sets. The green dotted line (``ORG'') is the original data tension error, while the blue dotted line is the
final data tension error used in the analysis, with uncertainty constrained to fall off at high $Q^2$ where the 
Mainz data do not contribute (see text for details). The purple dashed curves are the uncertainties associated
with the TPE corrections to the cross-section data at high-$Q^2$. The dashed black curves are the combinations of
these three sources of uncertainty, using the
data tension error that is cut off at high $Q^2$ (blue dotted line). The solid green curves are the parameterization
of the uncertainties provided in the Supplemental Material.}
\label{fig:protonuncertainties}
\end{center}
\end{figure*}


The proton fit includes 69 polarization extractions of $\gegm$, 
657 cross-section values~\cite{lee15} from the recent Mainz experiment~\cite{bernauer14}, 
and 562 cross-section values from other measurements, 
as well as the radius constraints and the high-$Q^2$ constraint points discussed above. 
The final fit yields a total $\chi^2$ of 1144.3 for 1306 degrees of freedom.
The $\gen$ ($\gmn$) fit includes 38 (33) data points, plus the radius and high-$Q^2$ constraints;
we obtain $\chi^2=24.50$ (29.56) for 45 (40) degrees of freedom.
It is not surprising that the reduced $\chi^2$ value is below unity for these fits: 
while the uncertainties quoted in the experiments are separated into scale uncertainties and uncorrelated contributions, 
in reality many of the systematic effects will have correlated contributions that vary with the kinematics in a nontrivial way.
Assigning uncorrelated uncertainties large enough to account for the unknown correlations in the data will
tend to yield lower $\chi^2$ values than one would expect for purely statistical or uncorrelated uncertainties.
In addition, for the bounded fit, each parameter adds both one degree of freedom and one constraint associated with
the Gaussian bound; thus, increasing the number of parameters does not reduce the number of degrees of
freedom, even though it does provide additional flexibility for the fit. Parameterizations of the fit central values and
uncertainties for all form factors are provided in the Supplemental Material~\cite{supplemental}.

Figure~\ref{fig:protonfits} shows the results of the fit for $\gep$ and $\gmp$ normalized to the dipole
form factor, $G_D = (1+ Q^2/\Lambda^2)^{-2}$ with $\Lambda^2=0.71\,{\rm GeV}^2$.
Points from a previous global analysis~\cite{arrington07c} of direct longitudinal-transverse (LT) separations for
$G_E^p$ and $G_M^p$ are also shown for comparison. Figure~\ref{fig:protongegm} shows the fit and uncertainties for
$\mugegm$ along with the direct extractions of $\mugegm$ from polarization measurements.


\subsection{Form Factors}

Figure~\ref{fig:protonuncertainties} shows the uncertainties for $\gep$ and $\gmp$ coming from
the covariance matrix of the fit, the systematic contributions accounting for the tension between 
different data sets, and the uncertainty associated with the TPE corrections at high $Q^2$.
Since the systematic contributions come from comparing two different fits 
(e.g., with and without the additional high-$Q^2$ TPE correction), the estimated corrections vanish
whenever the two fits cross. Such dips are artificial, and do not indicate a real reduction
in the uncertainties. For the TPE uncertainty, these dips occur only in regions where other
contributions dominate the uncertainties.  For the original data tension uncertainty (green dotted line labeled ``ORG''), these
dips yield an underestimate of the uncertainty for $Q^2$ values near 1~GeV$^2$, and it is necessary to
provide a better estimate of the uncertainty in this region. 
At high $Q^2$, the Mainz data only impact the fit through small normalization effects, 
and the green dotted line is driven by statistical fluctuations.
Because of these issues, we replace the dotted green line by a power law falloff after the first maximum 
(at around $Q^2 \approx 0.3$~GeV$^2$).
This fills in the artificial dips in the direct comparison of the fits,
and avoids letting the uncertainty grow at high $Q^2$ due to lack of data to constrain the fits.
The blue dotted line shows our final data tension error using the \textit{ad hoc} parameterization at higher $Q^2$.

The black dashed line is the combination of the various sources of uncertainty detailed above, and the
solid green line is a parameterization of this uncertainty, providing a simple closed form that provides
a good approximation at all $Q^2$ values. 
The parameterizations reproduce the complete uncertainty estimates with typical (RMS) deviations of $\sim 2\%$
except for $\gep$ in the $Q^2$ region from roughly 0.3--3~GeV$^2$. In this region, the total uncertainty is
dominated by our \textit{ad hoc} extension of the data tension uncertainty to higher $Q^2$, and as this is
the least rigorous part of the uncertainty extraction, we allow for larger deviations (typically a factor of 2--3)
in this region.


\begin{figure*}[ht]
\begin{center}
\includegraphics[angle=0,width=0.48\textwidth,height=8.0cm]{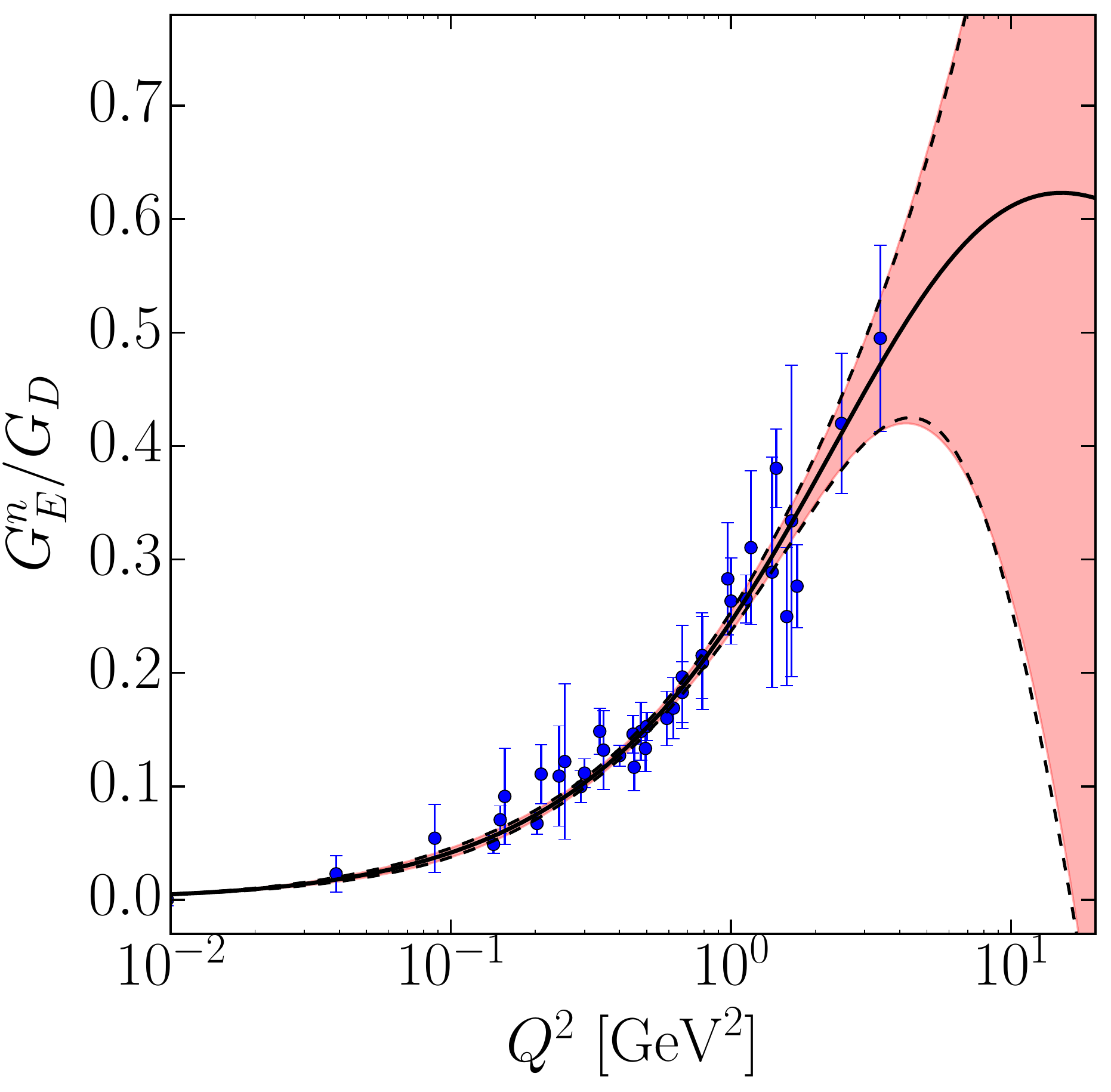}
\includegraphics[angle=0,width=0.48\textwidth,height=8.0cm]{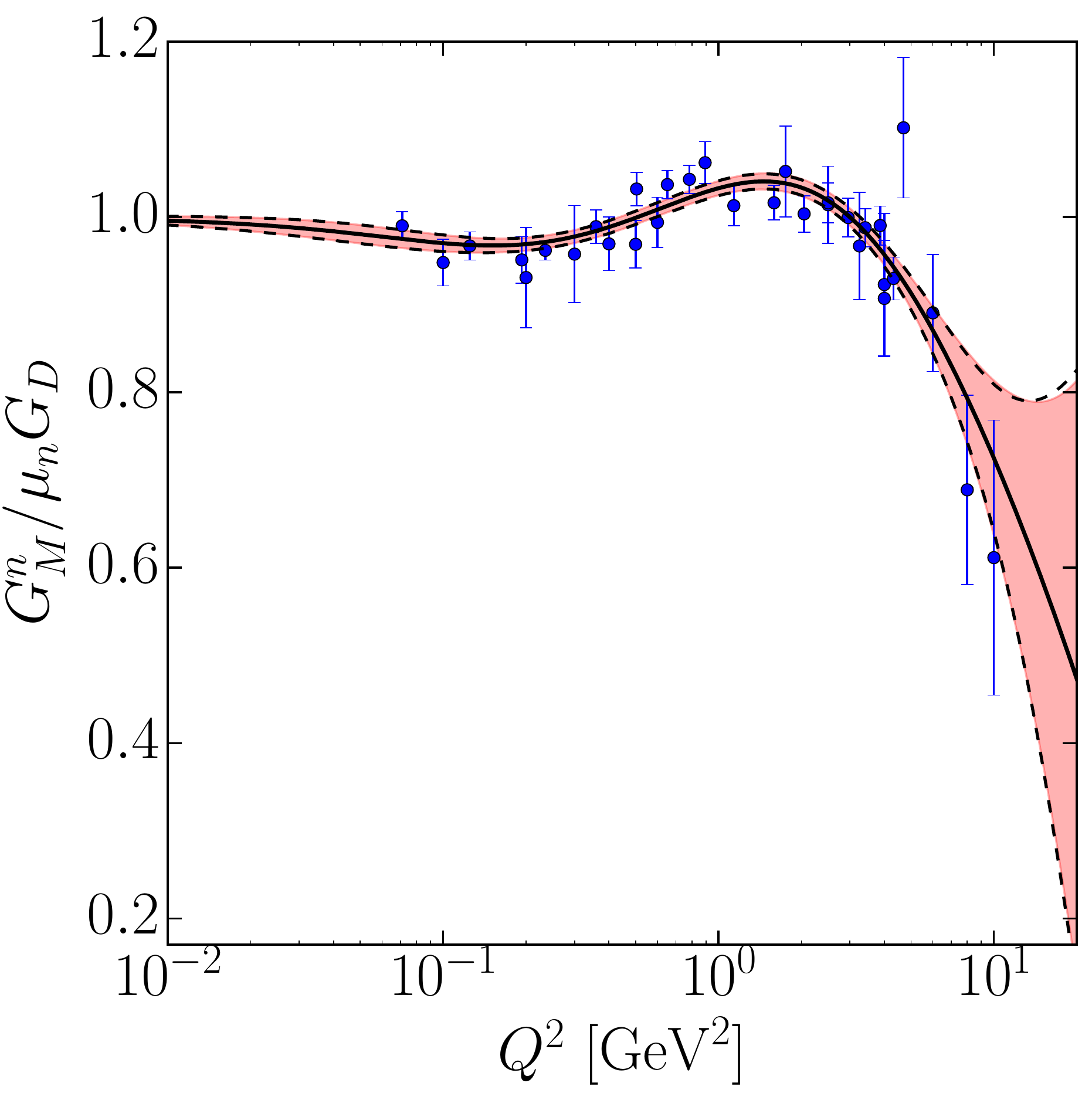}
\caption{(Color online) Parameterization of $\gen/G_D$ (left) (left) and $\gmn/\mu_n G_D$ (right) 
  from the global fit of neutron form factor data (solid curves). 
  The red shaded band is the fit uncertainty from the covariance matrix, and the dashed curves
  are the parameterization of the uncertainty provided in the Supplemental Material.
  The data points are the $\gen$ and $\gmn/\mu_n G_D$ values included in the fit.}
\label{fig:neutronfits}
\end{center}
\end{figure*}


Figure~\ref{fig:neutronfits} shows the fits to $\gen$ and $\gmn$, along with the data points
used in the fitting procedure. In this case, the uncertainties come from the
error matrix of the fit and represent the full uncertainties on the form factors;
tensions between different data sets have been accounted for in selecting the data for the
fit (as discussed earlier in Sec.~\ref{sec:ndata}). Calculations of the TPE corrections for the neutron~\cite{blunden05a, arrington11b}
yield smaller corrections than in the case of the proton, and we assume that the radiative correction uncertainties
already applied to the data are sufficient for the kinematics of existing data.


\subsection{Elastic ep cross sections \label{sec:elastic}}

The extracted form factors and uncertainties depicted in
Figs.~\ref{fig:protonfits}--\ref{fig:neutronfits} represent the
current state of knowledge for the nucleon electromagnetic form
factors, and are the primary result of this work.   They can be
applied to a range of precision observables.
For certain applications, including in legacy codes and in
experimental comparisons,  it is useful to work directly with the
elastic $ep$ cross sections instead of the form factors.
These cross sections can be
reconstructed from our representation of $\gep$ and $\gmp$, but  care
must be taken to reapply hard TPE effects in a fashion consistent with
the TPE correction applied to isolate the form factors studied in this
work: the hadronic calculations of Refs.~\cite{blunden05a, lee15},
plus the additional high-$Q^2$ correction of Eq.~(\ref{eq:extratpe}),
taken from Ref.~\cite{arrington07c}.
A complete reconstruction of the cross section would also account for
correlations in the errors of $\gep$ and $\gmp$. 

A practical alternative is to parameterize the cross section before subtracting the estimated TPE
corrections.  
We use the same fitting procedure as in our main analysis, excluding polarization data and neglecting
hard TPE corrections.
This provides a simple parameterization of the cross section that includes both
the Born and TPE contributions in ``effective" form factors.  
Note that we have not formally justified the $z$ expansion representation of the 
effective form factors, which now account for both one- and two-photon exchange processes.  
The effective form factor approach also enforces linear dependence of the reduced
cross section [i.e., the numerator in Eq.~(\ref{eqn:xs1photon})] on $\varepsilon$.
However, the TPE corrections are $\order(\alpha)$ and small, and detailed analyses
of world data~\cite{tvaskis06} show that $\varepsilon$ nonlinearities are also very small.
We do not pursue these questions in more detail here.

The effective form factors are not displayed here, but their central values are included in the
Supplemental Material~\cite{supplemental}.
The uncertainty associated with the TPE contribution in Fig.~\ref{fig:protonuncertainties}
should not be included in the effective form factor analysis since no hard TPE subtraction is being performed. 
However, this is never a dominant contribution to the cross section uncertainty. 
The $ep$ cross-section uncertainty is thus well approximated in the 
effective form factor approach by using the uncertainties from the main analysis, 
as displayed in Fig.~\ref{fig:protonuncertainties}.


\section{Summary}

We have performed global fits of electron scattering data to determine
the nucleon electromagnetic form factors and their uncertainties.
The form factor central values are presented as coefficients in the systematic
$z$ expansion framework, and error envelopes are also provided in
parameterized form. 
These form factors can be readily input to a range of precision observables. 

Our fits provide conservative and reliable errors that account for experimental tensions
and model uncertainties in the TPE corrections applied.  They are constrained in both low and high $Q^2$ limits,
with the goal of providing sensible extrapolations in both cases. 
At low $Q^2$, the fits have been constrained to consensus central values for
the nucleon charge radii and magnetic radii; as such, they do not provide new information on 
these quantities.
At high $Q^2$, power-law falloff has been enforced, consistent with the asymptotic scaling predictions of QCD;
however, the estimated uncertainties depend on theoretical priors and cannot be considered robust when
extrapolating beyond measured $Q^2$ values.

Our fit errors yield conservative uncertainty estimates 
compared to other analyses for specific applications and observables,
particularly those focused at low $Q^2$.   
This is due to the additional uncertainties we have assigned to 
account for tensions between different data sets.
These tensions can be further examined by selecting particular electron
scattering data sets or external radius constraints, 
in order to provide more precise predictions under different assumptions.
We will analyze some of
these observables in a future work~\cite{nextpaper17}.


\vskip 0.2in
\noindent
{\bf Acknowledgments.}
This work was supported by the U.S. Department of Energy, Office of Science, Office of Nuclear Physics 
under contract DE-AC02-06CH11357 and Office of High Energy Physics under contract
DE-FG02-13ER41958, and by a NIST Precision Measurement Grant. 
G.L. acknowledges support by the Israel Science Foundation (Grant No. 720/15),
by the United-States-Israel Binational Science Foundation (BSF) (Grant No. 2014397),
and by the ICORE Program of the Israel Planning and Budgeting Committee (Grant No. 1937/12).
G.L. acknowledges the hospitality of the Mainz Institute for Theoretical Physics, where part of this work was completed.
R.J.H. thanks TRIUMF for hospitality where a part of this work was performed.
Research at Perimeter Institute is supported by the Government of Canada through the Department of Innovation,
Science and Economic Development and by the Province of Ontario through the Ministry of Research and Innovation.
Fermilab is operated by Fermi Research Alliance, LLC under 
Contract No. DE-AC02-07CH11359 with the United States Department of Energy.

\bibliographystyle{elsarticle-num2}
\biboptions{sort&compress, numbers, comma, square}
\bibliography{globfit17}

\end{document}